\documentclass[aip,rsi,reprint]{revtex4-1}
\usepackage{graphicx}
\usepackage{epstopdf}
\usepackage{nicefrac}
\usepackage{setspace}

\begin{document}


\title{Cryogenic setup for trapped ion quantum computing} 




\author{M.F.~Brandl}
\email{Matthias.Brandl@uibk.ac.at}
\affiliation{Institut f\"ur Experimentalphysik, Universit\"at Innsbruck, Technikerstra{\ss}e 25, A-6020 Innsbruck, Austria}
\author{M.W.~van Mourik}
\affiliation{Institut f\"ur Experimentalphysik, Universit\"at Innsbruck, Technikerstra{\ss}e 25, A-6020 Innsbruck, Austria}
\author{L.~Postler}
\affiliation{Institut f\"ur Experimentalphysik, Universit\"at Innsbruck, Technikerstra{\ss}e 25, A-6020 Innsbruck, Austria}
\author{A.~Nolf}
\affiliation{Institut f\"ur Experimentalphysik, Universit\"at Innsbruck, Technikerstra{\ss}e 25, A-6020 Innsbruck, Austria}
\author{K.~Lakhmanskiy}
\affiliation{Institut f\"ur Experimentalphysik, Universit\"at Innsbruck, Technikerstra{\ss}e 25, A-6020 Innsbruck, Austria}
\author{R.R.~Paiva}
\affiliation{Institut f\"ur Experimentalphysik, Universit\"at Innsbruck, Technikerstra{\ss}e 25, A-6020 Innsbruck, Austria}
\affiliation{S\~ao Carlos' Physics Institute, University of S\~ao Paulo, Av. Trabalhador S\~ao-Carlense, 400, S\~ao Carlos, SP, Brazil}
\author{S.~M\"oller}
\affiliation{Dept. of Physics, University of California, Berkeley, CA 94720, USA}
\author{N.~Daniilidis}
\affiliation{Dept. of Physics, University of California, Berkeley, CA 94720, USA}
\author{H.~H\"affner}
\affiliation{Dept. of Physics, University of California, Berkeley, CA 94720, USA}
\author{V.~Kaushal}
\affiliation{QUANTUM, Institut f\"ur Physik, Universit\"at Mainz, Staudingerweg 7, 55128 Mainz, Germany}
\author{T.~Ruster}
\affiliation{QUANTUM, Institut f\"ur Physik, Universit\"at Mainz, Staudingerweg 7, 55128 Mainz, Germany}
\author{C.~Warschburger}
\affiliation{QUANTUM, Institut f\"ur Physik, Universit\"at Mainz, Staudingerweg 7, 55128 Mainz, Germany}
\author{H.~Kaufmann}
\affiliation{QUANTUM, Institut f\"ur Physik, Universit\"at Mainz, Staudingerweg 7, 55128 Mainz, Germany}
\author{U.G.~Poschinger}
\affiliation{QUANTUM, Institut f\"ur Physik, Universit\"at Mainz, Staudingerweg 7, 55128 Mainz, Germany}
\author{F.~Schmidt-Kaler}
\affiliation{QUANTUM, Institut f\"ur Physik, Universit\"at Mainz, Staudingerweg 7, 55128 Mainz, Germany}
\author{P.~Schindler}
\affiliation{Institut f\"ur Experimentalphysik, Universit\"at Innsbruck, Technikerstra{\ss}e 25, A-6020 Innsbruck, Austria}
\author{T.~Monz}
\affiliation{Institut f\"ur Experimentalphysik, Universit\"at Innsbruck, Technikerstra{\ss}e 25, A-6020 Innsbruck, Austria}
\author{R.~Blatt}
\affiliation{Institut f\"ur Experimentalphysik, Universit\"at Innsbruck, Technikerstra{\ss}e 25, A-6020 Innsbruck, Austria}
\affiliation{Institut f\"ur Quantenoptik und Quanteninformation der \"Osterreichischen Akademie der Wissenschaften,
Technikerstra{\ss}e 21a, A-6020 Innsbruck, Austria}


\date{\today}

\begin{abstract}
We report on the design of a
cryogenic setup for trapped ion quantum computing containing a segmented surface electrode trap.
The heat shield of our cryostat is designed to attenuate alternating magnetic field noise, resulting 
in 120~dB reduction of 50~Hz noise along the magnetic field axis.  
We combine this efficient magnetic shielding with high optical access required for single ion 
addressing as well as for efficient state detection by placing two lenses each with numerical aperture 0.23 inside the 
inner heat shield.  The cryostat design incorporates vibration isolation to avoid 
decoherence of optical qubits due to the motion of the cryostat.  We measure vibrations
of the cryostat of less than $\pm$20~nm over 2~s.  In addition to the cryogenic apparatus,
we describe the setup required for an operation with $^{\mathrm{40}}$Ca$^{\mathrm{+}}$ and
$^{\mathrm{88}}$Sr$^{\mathrm{+}}$ ions.  The instability of the laser manipulating the optical qubits in
$^{\mathrm{40}}$Ca$^{\mathrm{+}}$ is characterized yielding a minimum of its Allan 
deviation of 2.4$\cdot$10$^{\mathrm{-15}}$ at 0.33~s.  To evaluate the performance of the apparatus,
we trapped $^{\mathrm{40}}$Ca$^{\mathrm{+}}$ ions, obtaining a heating rate of 2.14(16)~phonons/s
and a Gaussian decay of the Ramsey contrast with a 1/e-time of 18.2(8)~ms.
\end{abstract}


\maketitle 

\section{Introduction}
\label{sec: Intro}

The applications of trapped ions in Paul traps range from
mass-spectrometry\cite{MassSpectrometry} and frequency
standards\cite{lineWidthCa,lineWidthSr} to quantum
simulation\cite{Cirac-Sim} and quantum computation (QC) where they enable
architectures which are promising candidates for fault-tolerant
QC\cite{Scalable_from_Kielpinski,KimFiberSwitch}.  There,
linear segmented Paul traps provide the means to
reconfigure the ion crystals where the quantum operations are applied
onto varying subsets of ions\cite{FirstPlanarTrap}.

Here, we describe an experimental setup containing a cryostat
as the cryogenic temperatures solve the following challenges towards 
large-scale QC with trapped ions.
(i)~Performing the necessary ion string reconfiguration operations
requires micro-fabricated traps which yield ion-to-electrode distances 
between 50 and 200~$\mathrm{\mu}$m. This vicinity of the ions to the electrode
surfaces results in a higher susceptibility to electric field noise
which disturbs the quantum operations\cite{heatingRatesMike}. It has
been observed that these detrimental effects are suppressed in traps that
are operated at cryogenic temperatures\cite{heatingRateOverTemp1,heatingRateOverTemp2}. 
(ii)~With increasing numbers of ions in the
traps\cite{quasi-Particle-engineering,quasi-Particle-engineering2},
collisions with background gas cause ion loss, making large scale QC impossible. 
Operating the trap in a cryogenic environment offers access to vacuum
pressures not accessible in room temperature setups.  A conservative
upper bound for vacuum pressures reached in cryostats is
10$^{\mathrm{-15}}$~mbar at a temperature of 4~K which is about 4 orders
of magnitude lower than typical vacuum pressures in room temperature
vacuum chambers\cite{CryogenicVacuum}. (iii)~When ions are used as a quantum memory,
magnetic field fluctuations are the main source of decoherence\cite{PhilippNJP}.  
This can be mitigated in a cryostat, as a
copper heat shield at cryogenic temperatures can be used to suppress 
alternating magnetic fields\cite{NISTmagnShielding}.

Operating an ion trap in a cryostat adds additional design challenges
as the operation of any type of cryostat causes vibrations in the
vacuum chamber. These vibrations have to be decoupled from the ion
trap, while maintaining a good thermal contact, as coherent manipulation
of optical qubits requires localization of the ions 
below the corresponding wavelength for the time of manipulation.

Furthermore, large-scale QC with trapped ions will involve ion crystal 
reconfiguration which will induce kinetic energy into the system\cite{bertha}.
To remove this energy, sympathetic cooling with multiple ion 
species\cite{SympatheticCooling,SympatheticCooling2} is employed
in our setup.  We focus on the
operation with $^{\mathrm{40}}$Ca$^{\mathrm{+}}$ and
$^{\mathrm{88}}$Sr$^{\mathrm{+}}$ ions and their respective optical
qubits\cite{qubitTypes} and describe the required laser
system and control electronics.

In Section~\ref{sec: cryogenic setup}, we describe our cryogenic setup 
and focus especially on vibrations, magnetic shielding against alternating 
magnetic fields, and the trap.  The laser setup required for our trapped ion QC
experiments is described in Section~\ref{sec: laser setup}.
Section~\ref{sec: electronics} covers the electronic setup used in
our experiments, and finally we present results of first characterization 
measurements with trapped ions in Section~\ref{sec: first experiments}.

\section{The cryogenic setup}
\label{sec: cryogenic setup}

When designing a cryogenic system, the first considerations address the type of cryostat.
For our setup, we considered closed cycle cryostats, like Gifford-McMahon and Pulse Tube
cryocoolers, which do not require a supply of liquid coolant, and wet
cryostats, like bath cryostats and (continuous) flow cryostats, which require a supply of liquid coolant
but do not require additional electrical power.
Wet cryostats do not produce acoustic noise, other than closed cycle cryostats. Such noise in the lab
will cause vibrations in all parts of the experiment.
When working with optical qubits\cite{qubitTypes} in a trapped ion QC setup, 
the experiments will be sensitive to the phase of the light 
at the position of the ion. Therefore, vibrations on the order of the wavelength of visible light 
reduce the coherence of an optical qubit. It is, thus, beneficial to work with a wet cryostat
which additionally allows tuning the temperature of the cold finger as well.
Hence, we decided to work with a flow cryostat\footnote{\textit{Janis} Model ST-400-1}.

Two heat shields, shown in Figure~\ref{fig: vibrational decouple}, are used to reduce the coolant consumption. 
The evaporated coolant from the inner stage cools the outer stage. 
Since the outer heat shield completely surrounds the inner heat shield, it will reduce the 
black-body-radiation induced heat load on the inner shield. 
Additionally, the heat load is further decreased by polishing and silver-plating all parts of 
the cryostat. 

\subsection{Mechanical stability and vibrations}
\label{ssec: stability and vibrations}

\begin{figure}[!tp]
	\begin{center}
		\includegraphics[scale=0.4]{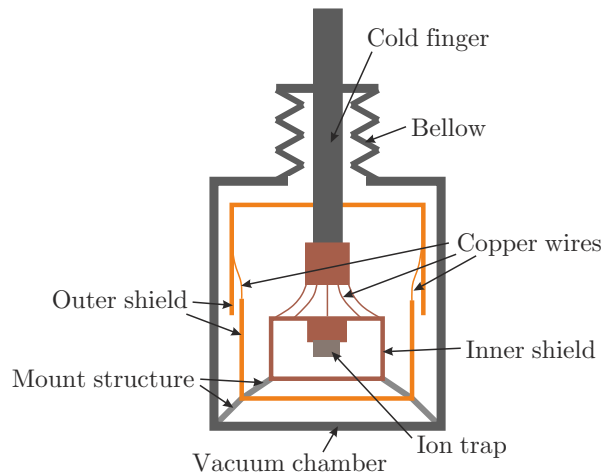}
	\end{center}
	\caption{Schematic of the vibrational decoupling. The inner shield, which
		contains the ion trap, is rigidly mounted to the vacuum chamber through mounts and
		the outer heat shield. The cold finger is connected via a bellow to the
		vacuum chamber and can vibrate independently of the vacuum chamber. The
		heat contact between the cold finger and the other heat shield is provided through
		thin copper wires.}
	\label{fig: vibrational decouple}
\end{figure}

Peak-to-peak vibrations due to the operation of a flow cryostat are usually on the
order of 1~{$\mathrm{\mu}$m} at the tip of the cold finger. \footnote{The actual vibrations can 
be significantly larger depending on how the cryogenic system is mounted.}
In order to suppress this movement, vibrational decoupling between the 
cold finger and the ion trap is required.  The incorporated decoupling 
scheme\cite{Tomaru} contains a membrane bellow which
reduces the mechanical coupling between the cold finger and the vacuum chamber,
as depicted in Fig.~\ref{fig: vibrational decouple}.
The inner shield, containing the ion trap, is mounted through rigid mounts with low 
thermal conductivity to the outer shield and the vacuum chamber.
At cryogenic temperatures, the cold finger and the heat shields are connected through thin oxygen-free high 
conducting (OFHC) copper wires, see Fig.~\ref{fig: vibrational decouple}, to reduce
the transmitted vibrations while retaining excellent thermal conductivity.  
Varying the number of wires enables finding a trade-off between 
vibrations and thermal contact.

\begin{figure}[!tp]
	\begin{center}
		\includegraphics[width=7cm]{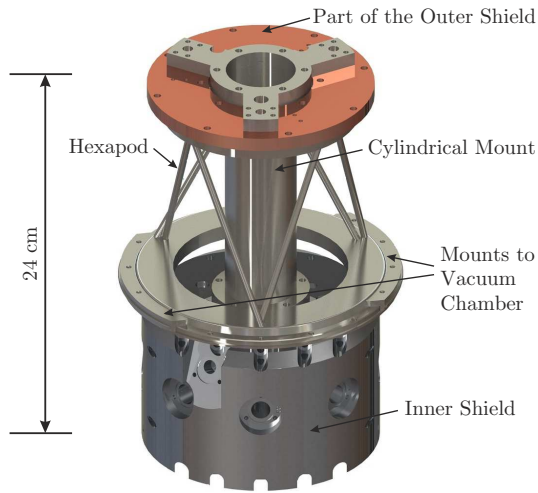}
	\end{center}
	\caption{Sketch of the rigid mounting structure. A hexapod connects
		the vacuum chamber and the outer shield, and a cylindrical mount connects
		the outer and the inner shield.}
	\label{fig: drawing of the cryostat mount structure}
\end{figure}

The mount structure connecting the outer shield to the vacuum chamber is a hexapod made of 316LN stainless steel,
as illustrated in Fig.~\ref{fig: drawing of the cryostat mount structure}.
Hexapods provide high stability at a small cross-section, thus combining high rigidity with
low thermal conductance. Due to space constraints, 
a stainless steel cylinder with a very thin wall ($\approx 0.5~mm$) connects the inner and 
the outer shield. This mount has a higher mechanical stability but also a higher 
thermal conductivity than a hexapod with the same cross-section. 
During operation with liquid helium, this is the largest contribution to the heat load of 
the inner heat shield, but still below 200~mW.

\begin{figure}[!bp]
	\begin{center}
		\includegraphics[width=7.5cm]{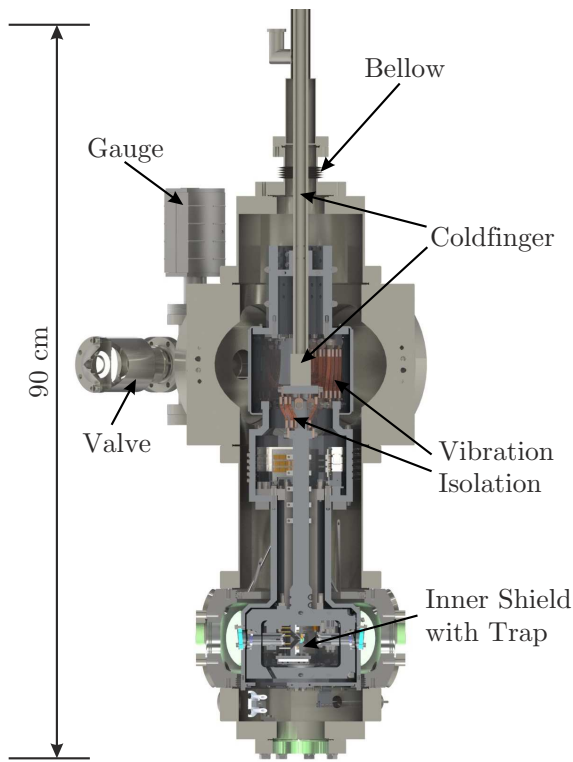}
	\end{center}
	\caption{A cutaway view of the cryostat.}
	\label{fig: cut through chamber}
\end{figure}

The whole vacuum chamber is illustrated as cutaway view in Figure~\ref{fig: cut through chamber}.
In addition to the vibration isolation in the vacuum vessel, aluminum cross bars
were installed outside the vacuum chamber to prevent movement of the vacuum chamber 
with respect to the optical table on which it is mounted.

The mount structure shown in Fig.~\ref{fig: drawing of the cryostat mount structure} is designed to
withstand force in the vertical direction, because it has to support about 23~kg of weight of
the copper shields.  The hexapod extends from the mount to the vacuum chamber to
the mount of the outer shield.  From there, the cylindrical mount connects to the lower
inner shield.  This mount structure resembles a double pendulum which is mechanically less 
rigid horizontally than vertically.  Hence, we only
investigated the vibrations in the horizontal plane, and assumed the vibrations in 
vertical direction to be much smaller than in the horizontal plane. 

\begin{figure}[!tp]
	\begin{center}
		\includegraphics[scale=0.5]{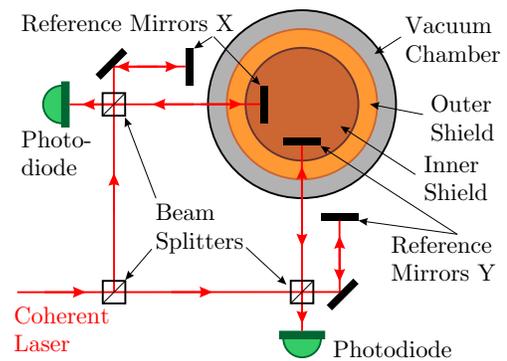}
	\end{center}
	\caption{Two Michelson interferometers employed to measure the vibrations
		in a 2D plane simultaneously.}
	\label{fig: vib measurement}
\end{figure}

An interferometric distance measurement with a Michelson interferometer was employed
to measure the differential length of the two interferometer arms, as depicted in Fig.~\ref{fig: vib measurement}. The two mirrors
of the interferometer arms were a reference mirror on the optical table and a mirror in 
the inner shield. This allowed us to measure the differential movement between the optical 
table and the inner shield. In order to measure the full movement of the inner shield, a 
second Michelson interferometer was placed perpendicular to the first one to record
the movements in two directions at the same time. 

\begin{figure}[!htb]
	\begin{center}
		\includegraphics[scale=0.35]{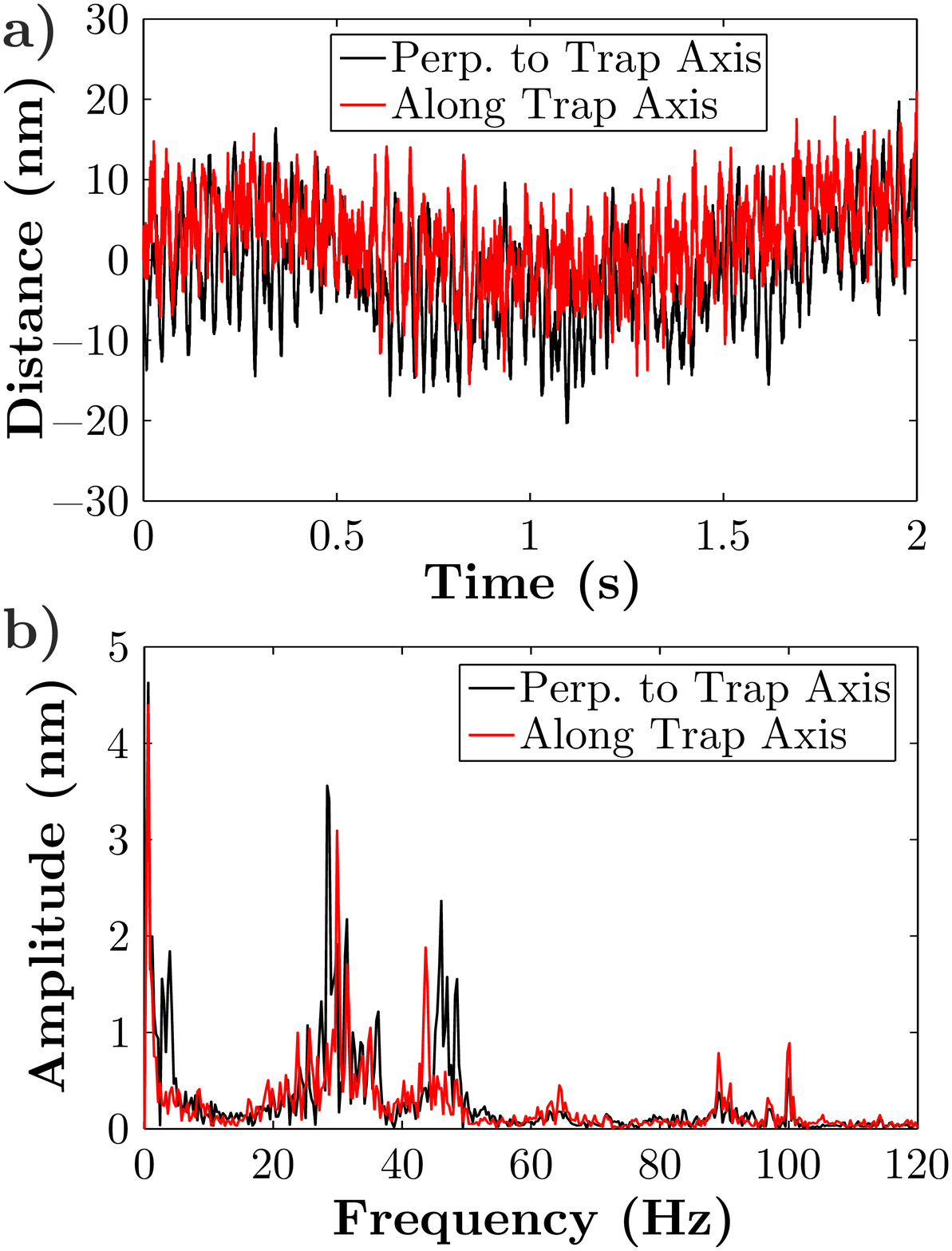}
	\end{center}
	\caption{The measured vibrations of the cryostat cooled with liquid nitrogen. Panel a) depicts the vibrations
		over time and Panel b) the Fourier-transform of these vibrations.}
	\label{fig: measured vibrations}
\end{figure}

The recorded movements are shown in Fig.~\ref{fig: measured vibrations}~a. 
The two measurement axes were chosen along and perpendicular to the axis of the
Paul trap in the cryostat. The vibrations along both axes are on the same order of magnitude
and stay between $\pm$20~nm over 2~s.  Fig.~\ref{fig: measured vibrations}~b depicts the 
Fourier transform of these vibrations. The peaks around 30 and 45~{Hz} correspond to the 
eigenfrequencies of the inner shield and the peaks between 90 and 100~{Hz}
to the eigenfrequencies of the vacuum chamber. 
The amplitude of the movement is much smaller than the wavelength 
of the qubit transitions, about 700~nm, allowing for coherent manipulation of optical qubits. 
All measurements were performed in absence of acoustic noise in the lab. Additional noise
can easily increase the vibration amplitudes by a factor of 10 or more.  

Long-term measurements over 10~min show an inner shield movement of less
than $\pm$120~{nm} from the original position.  We attribute this movement to the
vibration isolation of the optical table.  Due to small residual movements of the table, 
the vacuum chamber on the optical table is slightly tilted which 
leads to the observed movements. 
At the current state of trapped ion QC, typical quantum algorithms\cite{fourteen-qubit,DaniColorCode,MonzShor}
require a processing time of about 1~{ms}.  Further analysis showed that during 
3-5~{ms}, the phase shifts of the qubit laser light at the position of the trap
are dominated by acoustic vibrations coupling into the optical setup but
not by movement of the inner shield in the vacuum chamber.  Hence, our cryogenic
setup promises to have the same coherence properties as room temperature setups where
coherence times of tens of ms have been observed\cite{PhilippNJP}.

\subsection{Magnetic field shielding}
\label{ssec: magnetic field shielding}

Our setup is designed for operation with $^{\mathrm{40}}$Ca$^{\mathrm{+}}$ and $^{\mathrm{88}}$Sr$^{\mathrm{+}}$ ions. 
The available qubit transitions of these ion species have magnetic field dependences of up to 
2$\pi~\cdot$~39~{GHz/T}\cite{qubitTypes,magnDependence1,magnDependence2}. 
In order to not be limited by magnetic field fluctuations, the magnetic fields have
to be stabilized both spatially and temporally so that the frequency shift
due to magnetic perturbations is smaller than the natural linewidth of the qubit
transition, which for both ion species is the 
$^{\mathrm{2}}$S$_{\mathrm{1/2}}$~$\leftrightarrow$~$^{\mathrm{2}}$D$_{\mathrm{5/2}}$ transition.
In Ca$^{\mathrm{+}}$, the qubit transition has a linewidth of about 140~{mHz}\cite{lineWidthCa}, 
in Sr$^{\mathrm{+}}$ it is 400~{mHz}\cite{lineWidthSr}.
Hence, the magnetic field perturbations during one experimental cycle should not exceed 
$140~{\mathrm{mHz}}/\left(39~{\mathrm{GHz/T}}\right)~=~3.6~{\mathrm{pT}}$.
In our experiments, the typical quantization field strength is about 0.3~{mT}. Therefore, 
relative magnetic field stability better than about 1.2$\cdot$10$^{\mathrm{-8}}$ both
spatially and temporally is required. 

In this setup, the homogeneity is achieved by using Helmholtz coils,
which have an average radius of 19.5~{cm}.  Simulations show that this 
results in an inhomogeneity below the desired level of 1.2$\cdot$10$^{\mathrm{-8}}$ 
over a length of 1.5~mm around the trap center.  Hence, it will be possible to address ion 
strings over this length with a global beam simultaneously.  In order to achieve the desired 
magnetic field homogeneity, magnetic materials must not be used in the cryostat design. 

Technologically, it is challenging to actively stabilize 
magnetic fields with the required precision over time.  However, passive filtering
by exploiting the skin effect of a surrounding metal is feasible, as well as the generation of constant 
magnetic fields with superconducting coils.  
Since this setup is not big enough to incorporate superconducting Helmholtz 
coils with a radius of 19.5~{cm}, our choice was to reject alternating magnetic fields with 
thick OFHC copper walls\cite{NISTmagnShielding}.  The remaining dependence on 
slow drifts, e.g. set point of the current driver of the coils, or changes in earth's magnetic field, 
can be eliminated by interleaving actual measurements with calibration measurements\cite{ChwallaDiss}. 

The damping of alternating magnetic fields with OFHC copper relies on the
skin effect which prevents AC magnetic fields from penetrating a conductor. The skin
depth $\delta$ is given by
\begin{equation}
	\delta = \sqrt{\frac{2}{\omega \sigma \mu}}
	\label{eq: penetration depth}
\end{equation}
with $\omega$ the (angular) frequency of the magnetic field, $\sigma$  the 
conductivity of the material, and $\mu$ its permeability. At low temperatures, the 
conductivity of the copper becomes so high that magnetic field at frequencies as low 
as a couple of Hertz are suppressed over lengths in the cm or even the mm range. 

Experience has shown that the main spectral components of magnetic field fluctuations in 
trapped ion QC experiments are synchronous to the power line frequency.  Therefore, magnetic
field attenuation is most important at a frequency of 50~{Hz} and its harmonics.  
At room temperature, the skin depth of 50~{Hz} magnetic 
fields is about 9.2~{mm} in copper. The inner shield of our cryostat is made of ultra-pure copper that was 
thermally annealed in a nitrogen atmosphere at 450~{$^{\circ}$C} to increase its conductivity at cryogenic temperatures.
When cooling with liquid helium, we estimate a conductivity increase by a factor 
between 100 and 1000 as compared to room temperature.  Therefore, the skin depth due to skin effect at 50~{Hz} 
will decrease to a value between 0.92~{mm} and 0.29~{mm}. In this setup, the walls of the inner shield 
are 20~{mm} thick such that we calculate a magnetic shielding of -188~{dB} or even -600~{dB} for 50~{Hz} 
fields. These numbers elaborate that the cold, thick
copper walls are excellent magnetic shields for frequencies of 50~{Hz} and higher.  
In trapped ion experiments, the expected attenuation is lower due to holes in the shields
for optical access, and physical barriers for the eddy currents repelling the magnetic field fluctuations. For the latter case, contact resistances between
two parts of the shield are typically limiting the magnetic shielding. 

\begin{figure}[!tp]
	\begin{center}
		\includegraphics[width=7.5cm]{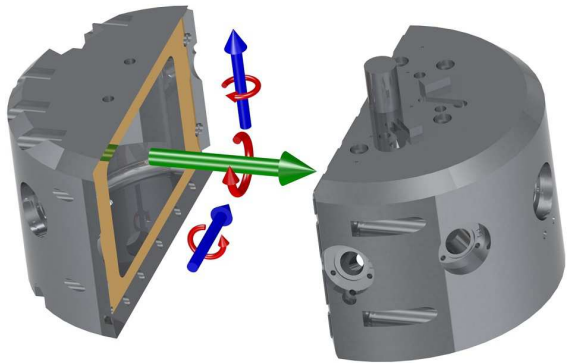}
	\end{center}
	\caption{Sketch of the two halves of the inner heat shield.  The green arrow symbolizes the
		magnetic field axis, the blue arrows represent the two axis perpendicular to the magnetic field axis,
		and the red arrows depicted induced eddy currents and their orientation.}
	\label{fig: eddy currents}
\end{figure}

For appropriate access to install the ion trap, its surrounding heat shield has to consist of at
least two parts. Eddy currents run in circles perpendicular to the magnetic field fluctuations that 
cause them. Therefore, the inner shield is cut 
into two pieces along a plane perpendicular to the quantization magnetic field, as can be seen in 
Fig.~\ref{fig: eddy currents}.   Along this axis (depicted by the green arrow), the sensitivity to magnetic fields is the strongest 
as fluctuations perpendicular to the axis have little influence on the absolute value of the magnetic field.
Thus, the shield was designed to have the strongest attenuation in this direction.  Eddy currents (depicted by red arrows) produced by magnetic fields along 
this axis run parallel to this plane and are not affected by the contact resistance between the two 
halves of the inner shield. Eddy currents caused by magnetic field fluctuations in different directions (depicted by blue arrows in Fig.~\ref{fig: eddy currents}) 
have to pass from one half to the other and will be limited by the contact resistance between the two 
halves of the heat shield at low temperatures. Therefore, the magnetic shielding will be less efficient in directions 
perpendicular to the quantization axis.

In order to characterize the magnetic shield, magnetic magnetoresistive sensors\footnote{\textit{Honeywell} 
HMC1001 and HMC1002} were placed inside the inner shield. 
We calibrated the sensitivity of the sensor for the current temperature by applying a step signal to the coils.

\begin{figure}[!tp]
	\begin{center}
		\includegraphics[scale=0.25]{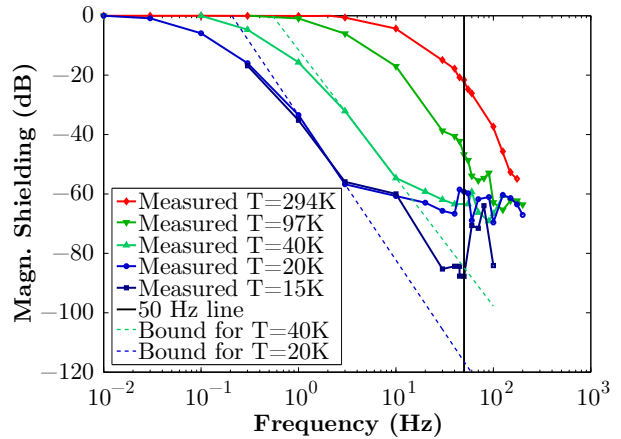}
	\end{center}
	\caption{The magnetic shielding over frequency along the quantization axis for various 
		temperatures of the inner heat shield.}
	\label{fig: magnetic shielding quantization axis}
\end{figure}

For the magnetic shielding measurement, a sinusoidal magnetic field of various frequencies was 
generated by the field coils and the response signal of the magnetic sensor was measured. 
Employing the calibration of the sensitivity, one obtains the attenuation for a specific frequency. 
The measurement results corresponding to the direction of the 
quantization axis can be seen in Fig.~\ref{fig: magnetic shielding quantization axis} for various temperatures. 
The measurement limit with our setup was between -60 and -55~{dB} and was mainly limited by 
pickup outside the vacuum chamber. 
The magnetic shielding improves with decreasing temperature.
Below 20~{K}, the attenuation saturates, which suggests that the electrical 
conductivity of the used copper is constant at temperatures of 20~{K} and lower.  
This means that in order to reduce the coolant consumption, it
is sufficient to operate the cryostat at a temperature of the inner shield slightly below 20~{K}.
Lower temperatures do not reduce the vacuum pressure, as hydrogen, 
the main constituent element of background gas, condensates at 20~{K}. 

\begin{table}[!bp]
\caption{Table of attenuations for 50~{Hz} magnetic fields at various temperatures. Extrapolated values are displayed in italics.}
\label{table: 50Hz attenuations}
\begin{tabular}{|l|c|c|c|c|}
	\hline
	\textbf{Temperature} & \textbf{294~{K}} & \textbf{97~{K}} & \textbf{40~{K}} & \textbf{20~{K}} \\
	\hline 
	\textbf{Attenuation} & 21~{dB} & 46~{dB} & \textit{ 85~{dB}} & \textit{120~{dB}} \\
	\hline 
\end{tabular}
\end{table}

Table~\ref{table: 50Hz attenuations} summarizes the measured attenuations of magnetic fields 
at 50~Hz and the estimated upper bounds.  For temperatures below the boiling point of liquid nitrogen, 
the attenuation for 50~{Hz} magnetic fields exceeds the measurement capabilities.  However, an upper 
bound for the attenuation can be determined.  The attenuation due to skin effect along a direction $x$ 
scales as $\exp\left(-\nicefrac{x}{\delta}\right)$ with the skin depth $\delta$ from 
eq.~\ref{eq: penetration depth}.  Hence, the gradient of the attenuation becomes greater with 
increasing frequency, if the attenuation is limited by skin-effect. 
Extrapolating the last two data points above this noise floor, we estimate an attenuation of 120~dB as 
indicated in Fig.~\ref{fig: magnetic shielding quantization axis}. 

\begin{figure}[!tp]
	\begin{center}
		\includegraphics[scale=0.25]{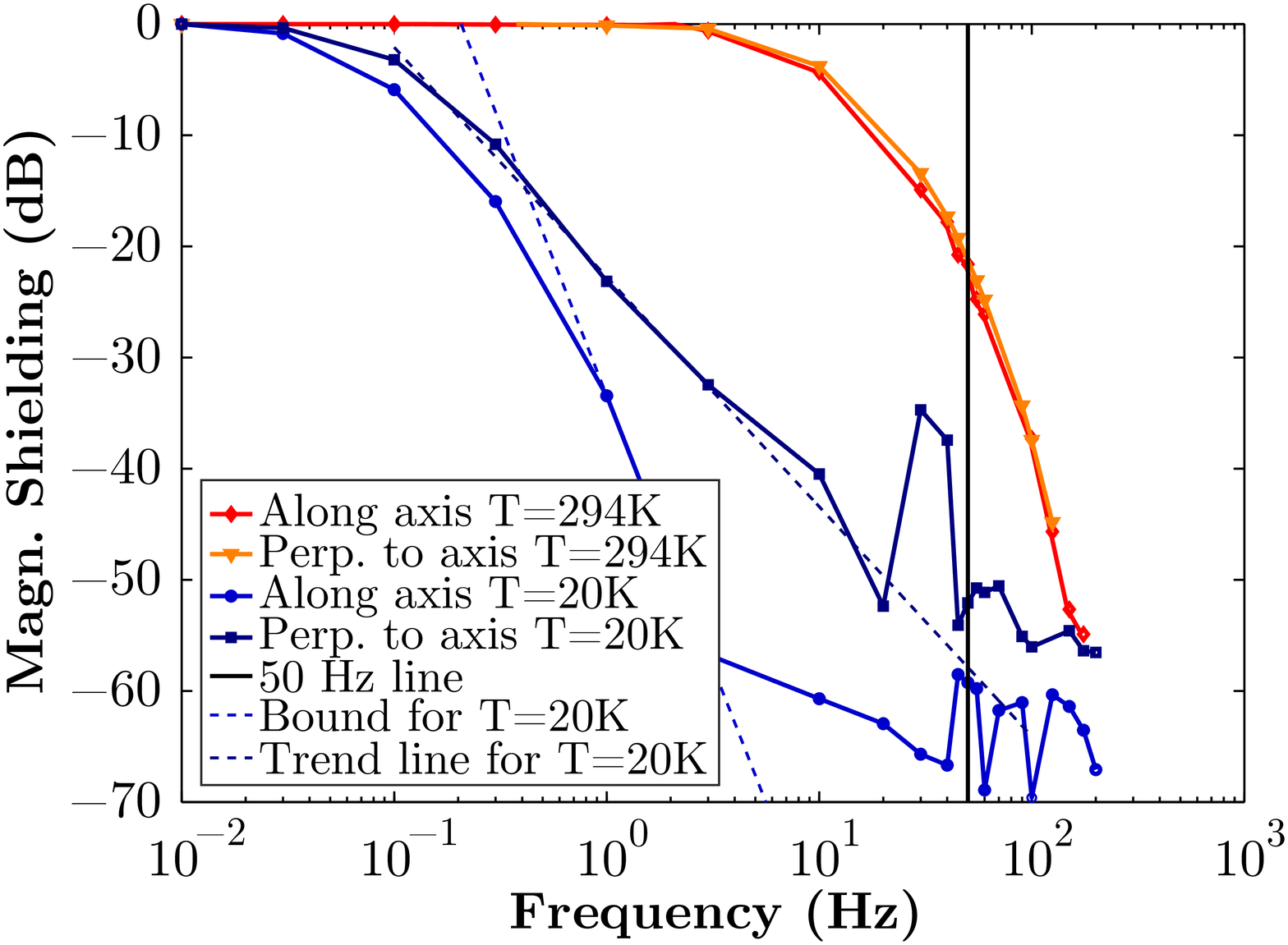}
	\end{center}
	\caption{The magnetic shielding along the quantization axis and perpendicular
		to it, when the inner heat shield is at room temperature and at 20K.}
	\label{fig: comparison magn shielding HH and Hor}
\end{figure}

The comparison of the magnetic shielding along the quantization axis and perpendicular to it is depicted in 
Fig.~\ref{fig: comparison magn shielding HH and Hor}. 
At room temperature, the contact resistance is not the limiting factor, and there is no significant difference 
between the magnetic shielding along and perpendicular to this magnetic 
field axis. By contrast at low temperatures, the conductivity of the copper increases sharply, and the contact resistance 
becomes the limiting factor. Hence, the magnetic shielding perpendicular to the axis will be worse than along 
the axis. Since the contact resistance is not frequency dependent, magnetic shielding limited by contact 
resistance will have a constant gradient in a double logarithmic plot. 
As shown in Fig.~\ref{fig: comparison magn shielding HH and Hor}, 
the shielding of magnetic fields of 50~{Hz} perpendicular to the quantization
axis is about -60~{dB}, whereas along the axis the upper bound is -120~{dB}.

The measured attenuation was so high that the trapped ion qubits are 
decoupled from ambient alternating magnetic fields which are typically in the low $\mathrm{\mu}$T regime 
in typical laboratory environments.  Slow magnetic field drifts, like offset drifts of the current in the coils 
or changes in earth's magnetic field, will still affect the qubits in the inner shield.
However, dynamical decoupling sequences\cite{DynamicalDecoupling} with a periodicity faster 
than the frequencies that can penetrate the inner shield decouple the experiments from these low frequency magnetic field drifts. 
Additional calibration measurements between quantum information experiments can be introduced to
infer the qubit transition frequencies\cite{ChwallaDiss}.

\subsection{Trap and beam geometries}
\label{ssec: trap and beam geometries}

\begin{figure*}[!htb]
	\begin{center}
		\includegraphics[width=15.5cm]{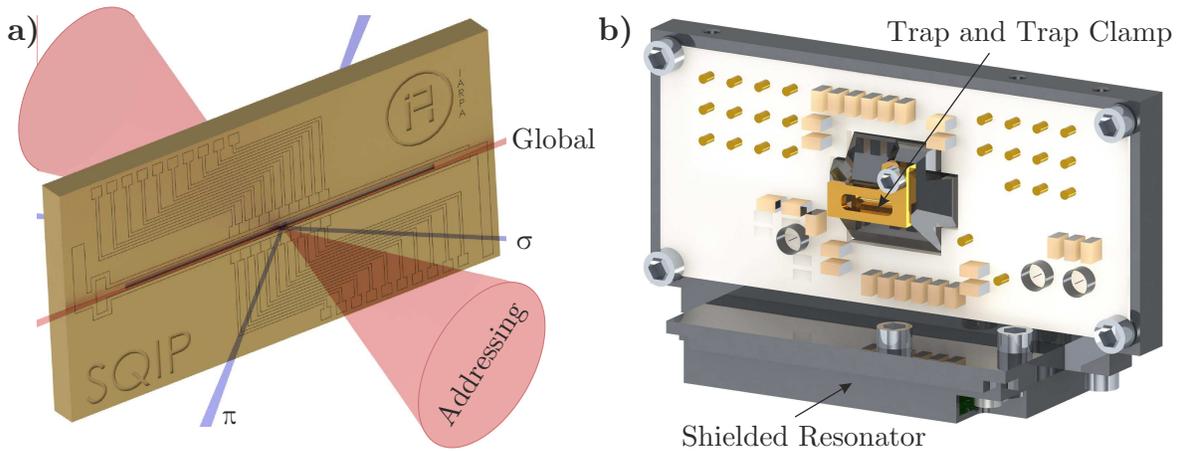}
	\end{center}
	\caption{Panel a) shows a sketch of our trap with the laser beams for state manipulation.
		Panel b) displays a the trap mounting structure with the attached shielded resonator. }
	\label{fig: trap pic}
\end{figure*}

A universal set of quantum gates for quantum computing\cite{fullSetQGates} requires the capability 
to address single ions\cite{AddressingHCN}.  In order to optically address single ions with a wavelength 
of about 700~nm and typical axial trap frequencies of up to 1~{MHz}\cite{LevelSchemes}, 
a diffraction-limited addressing requires a numerical aperture (NA) of about 0.2 or higher.  
Following the approach of reference \onlinecite{Scalable_from_Kielpinski}, a scalable trapped ion 
quantum computer will require a structured Paul trap.  
For ease of scalable trap fabrication, our trap is designed such that all electrodes lie in a single plane, 
as illustrated in Fig.~\ref{fig: trap pic}~a.  Two RF-electrodes, both 60~$\mathrm{\mu}$m wide, provide radial confinement. 
The outer DC segments have a width of 200~$\mathrm{\mu}$m separated by 10~$\mathrm{\mu}$m wide and 30~$\mu$m 
deep gaps from the other electrodes.  The 100~$\mathrm{\mu}$m wide slot in the center electrode allows for light 
with a numerical aperture of up to 0.24 to pass through without scattering from the edges of the slot. 
The trap was structured by \textit{Translume}\footnote{Translume Inc., 655 Phoenix Dr, Ann Arbor, MI 48108, 
United States} and gold-coated in-house by angle evaporation\footnote{In total, we used four different angles 
for evaporation to prevent electric connection through the trenches, while making sure that the side walls as 
well as the center slit are coated.  Under each angle we first deposited 10~nm of titanium for adhesion and 
then 300~nm of gold.} to prevent electric connection through the trenches and to cover the central slit with metal.

In typical experiments with segmented ion traps, all beams are parallel to 
the plane of the trap.  The required NA for single-ion addressing makes different beam geometries
necessary.  As depicted in Fig.~\ref{fig: trap pic}~a, our trap has a slit through which all beams
have to pass.  The only exception is a global qubit beam, which runs parallel to the surface and addresses
all ions in the trap simultaneously.  The addressing qubit beam with the high NA for single-ion addressing 
is perpendicular to the surface. The beams used for
cooling, detection and initialization, depicted in blue, are oriented 45$^{\circ}$ to the trap surface.
In order to allow state initialization with optical pumping, one of these two beams is oriented
along the magnetic field axis of the experiment ($\mathrm{\sigma}$-light). The other one is perpendicular to it
($\mathrm{\pi}$-light).

Electrostatic simulations predict a pseudo-potential minimum of 100~$\mathrm{\mu}$m above the trap surface, 
resulting in a minimal electrode-to-ion distance of 113~$\mathrm{\mu}$m.  In order to define the boundary 
conditions more reliably, a gold-plated clamp displayed in Fig.~\ref{fig: trap pic}~b is placed around the trap. 
In addition, the central electrode is connected to an RF connection to allow for global qubit rotations between 
Zeeman substates using RF-fields.

\begin{figure}[!bp]
	\begin{center}
		\includegraphics[width=7cm]{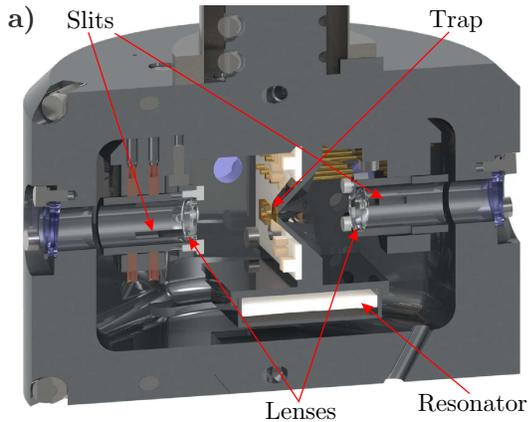}
	\end{center}
	\caption{Trap and optical assembly inside the inner shield. Panel a) depicts a cut through the 
		sketch of the inner heat shield with the trap, the resonator and the lenses for addressing 
		and detection. Panel b) shows a picture of a lens mount.}
	\label{fig: inner shield}
\end{figure}

If the addressing and detection optics were placed entirely outside the cryostat, the desired NA of 0.2
would require holes of large diameter in the heat shield, which would deteriorate the magnetic shielding efficiency. 
In order to achieve the required NA, we placed lenses
close to the ion trap in the cryogenic environment, such that only collimated beams of small diameter 
have to pass through the heat shields.  To increase the collection efficiency of scattered photons, two lenses,
one at the front and one at the back of the trap, were placed inside the inner shield, as displayed
in Fig.~\ref{fig: inner shield}~a.

The used aspherical lenses with half-inch diameter\footnote{\textit{Thorlabs} AL-1225} have a focal length of 25~{mm} and 
an NA of 0.23.  Their custom anti-reflection-coating allows using the lenses for single ion addressing 
with the qubit lasers around a wavelength of 700~{nm} as well as for detection of fluorescence around a wavelength of 400~{nm}. 
Thermal contraction of the lens and its mount leads to mechanical stress on the lens which will 
cause birefringence in the lens.  To circumvent this, a cylindrical mount with long slits was designed, shown 
in Fig.~\ref{fig: inner shield}~b.  The long slits act as flat springs to reduce mechanical stress.
Because of the round shape of this mount, the remaining mechanical stress should be symmetrical, and
thus not cause birefringence.  

An advantage of having the lenses in the inner shield is that
the addressing beams can pass through the glass of the viewport, the outer and inner shield as 
collimated beams. That way, aberrations deteriorating the qubit addressing capabilities are reduced.
Alternatively, complicated objectives consisting of several lenses outside the vacuum would need to 
compensate for aberrations caused by the viewport.
Typically, these are more sensitive to aberrations caused by off-center or tilted beams than a single lens.
Furthermore, as the ion trap, the inner shield and the focusing lenses represent one rigid mechanical 
assembly, alignment errors due to small movement of the inner shield are strongly suppressed 
compared to external focusing optics.  The only significant misalignment in the path of the 
addressing beam can originate in a change of the angle under which the incoming beam hits the last lens. 
This is caused outside of the vacuum vessel and has to be considered in the design of the addressing optics.

Using the lenses for detection, each lens collects about 1.3~\% of the light emitted by
the ions in the focal point of the lens.  With a typical magnification of about 30,
we are able to image ion strings of 50 ions onto our electron multiplying charge-coupled 
device (EMCCD) camera\footnote{Andor iXon+ EMCCD-Detector, 512 $\times$ 512 pixel, 16 $\mu$m}.  

The resonator to generate the RF voltage in the Paul trap was a shielded RLC-series
resonator.  It is mounted underneath the trap in the inner heat shield.  
The resonator coil is made of a high temperature superconductor
with a critical temperature above 87~K.  It has an inductance of 1.6~$\mu$H resulting in a 
resonance at 49.9~MHz.  Details on the design of this resonator can be found in reference \onlinecite{cryoResonator}.
With less than 100~mW of RF power, we could load ions and operate the trap.

\section{Laser setup}
\label{sec: laser setup}

\begin{figure}[!htb]
	\begin{center}
		\includegraphics[scale=0.27]{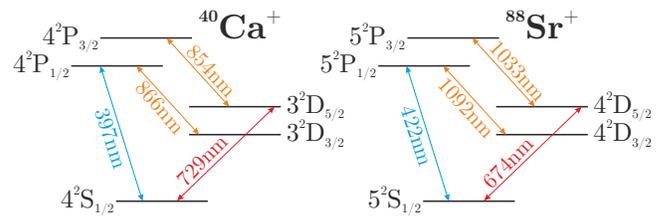}
	\end{center}
	\caption{Energy level diagrams for $^{40}$Ca$^+$ and $^{88}$Sr$^+$}
	\label{fig: level schemes}
\end{figure}

\begin{figure*}[!htb]
	\begin{center}
		\includegraphics[width=16cm]{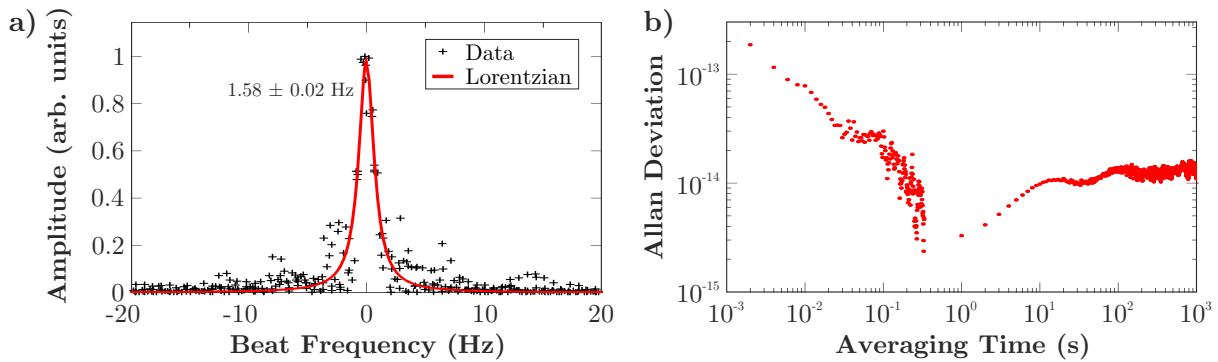}
	\end{center}
	\caption{Panel a) illustrates the beat note of our 729nm-laser with a second 729nm-laser.
		The Lorentzian fit results in a linewidth of 1.58(2)Hz. Panel b) shows the Allan deviation over averaging time. 
		The lowest observed Allan deviation was 2.4 $\cdot$ 10$^{\mathrm{-15}}$ at an averaging time of 0.33~{s}.}
	\label{fig: allan deviation}
\end{figure*}

Motional heating\cite{heatingRatesMike} poses a limit on quantum information processing in large 
scale ion trap quantum computers, and it will require recooling of ion strings without affecting 
the quantum states of the qubits.  A suitable technique is sympathetic 
cooling\cite{SympatheticCooling,SympatheticCooling2}, where a second ion species is used to
cool heterogeneous ion crystals.
Therefore, we work with $^{\mathrm{40}}$Ca$^{\mathrm{+}}$ and $^{\mathrm{88}}$Sr$^{\mathrm{+}}$ ions.
Their level structure, which can be seen in Fig.~\ref{fig: level schemes}, is identical, while the 
corresponding transition wavelengths and radiative lifetimes are different.  Henceforth, when we denote 
transitions without specifying an ion species, it means that the statement is applicable to both ion species.  
In our approach to trapped ion QC, read-out and quantum state manipulation are done with 
timed light pulses.  The optical qubit is encoded in the transition 
$^{\mathrm{2}}$S$_{\mathrm{1/2}}$~$\leftrightarrow$~$^{\mathrm{2}}$D$_{\mathrm{5/2}}$. 
This is a dipole forbidden transition, and hence, $^{\mathrm{2}}$D$_{\mathrm{5/2}}$ is a 
metastable state with a life-time on the order of 1~{s}.  The D-state life-time of each species is long compared to 
typical gate times on the order of 10~{${\mathrm{\mu}}$s} and allows coherent manipulation.  
Coherent optical driving of the transition requires a constant phase of the light, which is 
only ensured with a laser whose frequency is constant during the qubit manipulation. 
State detection is a projective quantum measurement for which we 
employ state-dependent resonance fluorescence.  There, laser beams resonant to the transitions 
$^{\mathrm{2}}$S$_{\mathrm{1/2}}$~$\leftrightarrow$~$^{\mathrm{2}}$P$_{\mathrm{1/2}}$, and
$^{\mathrm{2}}$D$_{\mathrm{3/2}}$~$\leftrightarrow$~$^{\mathrm{2}}$P$_{\mathrm{1/2}}$ will 
induce fluorescence of an ion if the outer electron is in the $^{\mathrm{2}}$S$_{\mathrm{1/2}}$ 
state.  If the outer electron is in the $^{\mathrm{2}}$D$_{\mathrm{5/2}}$ state, the ion will 
stay dark.   To reset the qubit, an additional laser for the transition 
$^{\mathrm{2}}$D$_{\mathrm{5/2}}$ $\leftrightarrow $$^{\mathrm{2}}$P$_{\mathrm{3/2}}$ is employed.
The exact wavelengths for these transitions can be found in references \onlinecite{LevelSchemes,lineWidthCa,lineWidthSr}. 

\begin{figure*}[!htb]
	\begin{center}
		\includegraphics[width=15cm]{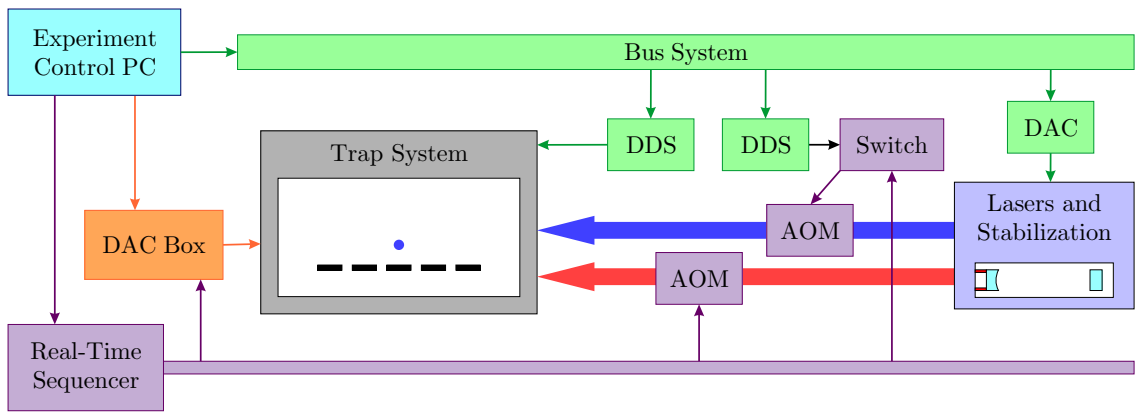}
	\end{center}
	\caption{The schematic of the experiment control. Before the measurement, the preset signals like the voltages to set
		the laser frequencies or the trap drive RF are programmed via the bus system. During the experiment,
		the real-time sequencer switches qubit lasers directly on and off with internal DDSs. Other laser beams are
		controlled via RF switches. Furthermore, the sequencer triggers the DAC box to control the trap DC 
		voltages of the trap.}
	\label{fig: experiment control}
\end{figure*}

In the following, we will discuss the lasers required for the operation of our experiment.
To load ions into the trap, an oven is heated up until it emits a beam of neutral atoms. 
In this setup, there are two commercial ovens\footnote{\textit{Alvatec}, Gewerbestra{\ss}e 3, 9112 Griffen, Austria}, 
one filled with Ca and one filled with Sr, mounted inside the vacuum chamber without thermal contact to the cryostat.  
These atoms are ionized by photo-ionization\cite{PI-Ca,PI-Sr}. 
For the isotope-sensitive laser excitation during the ionization process, 
external cavity diode lasers (ECDL) are used\footnote{\textit{Toptica} DL Pro from Toptica Photonics AG,
Lochhamer Schlag 19, 82166 Gr{\"a}felfing, Germany}.  Broadband light sources
with enough intensity in the required wavelength range are employed for the non-isotope-sensitive
part of the ionization.

The lasers driving the dipole transitions in both ion species are ECDLs\footnote{\textit{Toptica} 
DL pro}, which are frequency stabilized to reference cavities using the Pound-Drever-Hall (PDH) 
scheme\cite{PDH}. The sole exception is the 397~{nm} laser\footnote{\textit{Toptica} TA SHG pro}
which is a frequency-doubled laser and the light at 794~{nm} is locked to a cavity.
The typical finesses of the reference cavities are between 500 and 1000, and the free 
spectral range (FSR) is 1.5~{GHz}.  This results in typical linewidths on the order of 
100~{kHz} for our lasers.  One of the two mirrors of each cavity is mounted on piezos 
which allows very precise frequency tuning.  Hence, it is possible to tune each cavity to 
an arbitrary frequency around the corresponding dipole transition of the ion.

In our experiments, the quadrupole $^{\mathrm{2}}$S$_{\mathrm{1/2}}$~$\leftrightarrow$~$^{\mathrm{2}}$D$_{\mathrm{5/2}}$ 
transition is used as the qubit transition\cite{qubitTypes,LevelSchemes,lineWidthCa,lineWidthSr}. 
To achieve coherence times longer than 100~ms, a linewidth of the laser driving this transition 
of less than 10~{Hz} is required. Both qubit lasers are ECDLs\footnote{\textit{Toptica} TA pro} with an 
extended optical resonator.  The free running natural linewidth of such an ECDL can be on the order of 1~{MHz}.
Hence, the frequency stabilization has to reduce the linewidth by five or six orders of magnitude. 
This is challenging with a single stage, and thus, a two-stage approach\cite{multiStageLock} 
was pursued for the 729nm laser and the 674nm laser. 

The first stage is a PDH lock to a pre-stabilization cavity with medium finesse (MF) of about 10000, 
which is located on the same cavity spacer as the optical resonators for the dipole lasers, 
yielding a FSR of 1.5~{GHz}.  This results in a linewidth between 1 and 10~{kHz} after the 
first locking stage. 

The second stage is another PDH lock in which the feedback acts through a voltage controlled 
oscillator (VCO) driving an acousto-optic modulator (AOM). This feedback loop is referenced to a
high-finesse (HF) cavity\footnote{\textit{Stable Laser Systems}, 4946 
63rd Street, Suite B, Boulder, CO 80301, USA} with a finesse of about 250000 with 
and an FSR of 3~{GHz}. This light is then amplified with a tapered amplifier 
before it is sent to the experiment. 

Since the pre-stabilization cavity contains two piezos, which have a non-zero thermal expansion 
(of first order), its length will thermally drift with respect to the HF cavity, making their 
resonance frequencies drift apart.  Due to the limited bandwidth of the VCO driving the AOM,
this drift is compensated with a field programmable gate array (FPGA)-based counter which feeds 
back on one of the piezos of the MF cavity to compensate the thermal drift.  

In order to measure the linewidth of the laser, beat measurements of the 729nm laser with the laser 
used in reference \onlinecite{lineWidthCa} were performed.  The results of these beat measurements 
can be seen in Fig.~\ref{fig: allan deviation}. With a finesse of 242742(6) of our 729nm cavity, a 
Lorentzian fit of the beat signal of the two 729nm lasers resulted in a linewidth of 1.58(2)~Hz\cite{MasterPostler}.  
Additionally, the Allan deviation of the beat signal was calculated and resulted in a lowest 
Allan deviation of 2.4$\cdot$10$^{\mathrm{-15}}$ at an averaging time of 0.33~{s}.  
We expect a similar performance from our 674nm laser as we use the same lock system for both lasers. 
However, due to the lack of a second 674nm laser of similar linewidth, we could not perform the same 
measurements with our 674nm laser.

\section{Electronics}
\label{sec: electronics}

In trapped ion QC experiments, the following signals have to be accurately controlled:
\begin{itemize}
	\item Radio-frequency (RF) for the RF-drive of the Paul trap
	\item DC voltages for the cavity piezos to tune the laser frequencies
	\item Light has to be switched on and off with sub-microsecond precision
	\item Signals for phase-coherent qubit manipulation
	\item DC voltages to confine the ion string(s) axially
\end{itemize}
It is beneficial to distinguish between quasi-static signals and dynamical fast signals.
Quasi-static signals are signals which can be set before a measurement.  They include
for example the amplitude and frequency of the RF voltage to drive the Paul trap 
or the DC voltages applied to cavity piezos to tune the laser frequencies.  The dynamical 
fast signals have to be varied during a measurement.  They contain for example the phases
of RF signals required for phase-coherent manipulation of laser light.

To control the quasi-static signals, we use a digital bus system\cite{SchreckBusSystem} developed at the University of Innsbruck, 
as depicted in Fig.~\ref{fig: experiment control}. The bus has 8 address
bits, 16 data bits, and 1 strobe bit, which are sent via a 50 pin ribbon cable, where every second pin is
grounded.  The ground lines in the ribbon cable cause a considerable capacitance, which limits the transmission length
to about 5~{m} before the signal has to be digitally refreshed to provide a bit error probability below 10$^\mathrm{-8}$ at
a bus timing of 1~{${\mathrm{\mu}}$s}. The digital refreshing is performed by bus driver cards, which 
provide galvanic isolation between their input signals and their output signals and, thus, also prevent
ground loops.

The bus system has output cards for analog voltages, digital voltages, and RF signals.  These output cards
allow us to generate the signal close to their targets. 
The RF signals are generated with direct digital synthesizer (DDS) chips, and allow setting an 
RF signal before a measurement sequence, which can be switched on and off via an RF switch during a 
sequence. The switching of an RF signal can be converted to switching of an optical signal with an 
AOM in single pass or double pass configuration.
The output cards are mounted in rack systems, which additionally support RF amplifiers, sample-and-hold 
PID controllers, and other cards that are of use in the lab. 

In trapped ion QC experiments, typical single qubit gate times are on the order of 1 to 10~{${\mathrm{\mu}}$}s. 
When all quasi-static signals are set, our real-time sequencer with a timing resolution of 10~{ns} starts
the measurement. The sequencer, depicted in Fig.~\ref{fig: experiment control}, can control digital 
outputs and react to digital inputs.  Furthermore, it features DDSs, which allow phase-coherent frequency 
switching and amplitude shaping of RF signals.

The DC voltages for confining and moving ions are supplied by an FPGA-based arbitrary waveform generator 
designed and built at the University of Mainz\cite{bertha}. This arbitrary waveform generator 
is triggered by a digital signal from the real-time sequencer to allow movement of ions during a measurement.

\section{Experiments with trapped ions}
\label{sec: first experiments}

Two non-evaporative getters\footnote{\textit{SAES Getters}, Viale Italia, 77, 20020 Lainate MI, Italy} in the
vacuum chamber allow us to typically achieve pressures of below
10$^{\mathrm{-8}}$~mbar at room temperature. Then, we precool for 24 to
36~hours with liquid nitrogen to reduce the liquid helium consumption during further cool down. After switching
to liquid helium, it usually takes a couple of hours to reach our typical operating temperatures of about
20~K.  At these temperatures, our cryogenic system typically consumes 0.5~l liquid helium per hour.  We estimate 
that only 50\% of this consumption is due to thermal leakage in the experimental vacuum chamber.  The remaining
50\% of this consumption are caused by the dewar and the transfer line.

\begin{figure}[!bp]
	\begin{center}
		\includegraphics[scale=0.3]{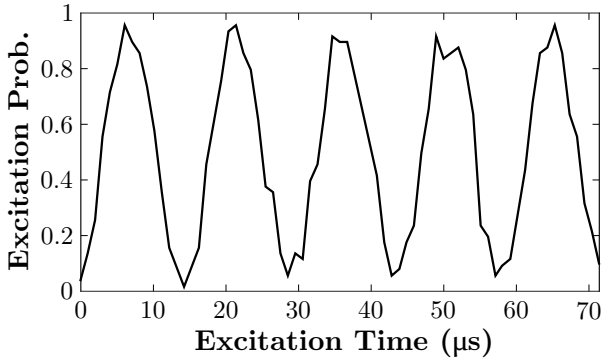}
	\end{center}
	\caption{Rabi flops with a sideband cooled ion.}
	\label{fig: rabi carrier}
\end{figure}

Under typical operating temperatures of about 20~K, 
we trap ions and obtain trap lifetimes of several hours under continuous Doppler cooling.  
We recorded Rabi flops on the carrier with a 
sideband cooled $^{\mathrm{40}}$Ca$^{\mathrm{+}}$ ion using the global beam, shown in 
Fig.~\ref{fig: rabi carrier}.  A measurement of the average phonon number resulted
in roughly 0.1 phonons along the trap axis, whereas the observed decay corresponds to
an average phonon number of 14.  Therefore, we assume that the ion was radially at a temperature
well above the Doppler cooling limit.

\begin{figure}[!tp]
	\begin{center}
		\includegraphics[scale=0.3]{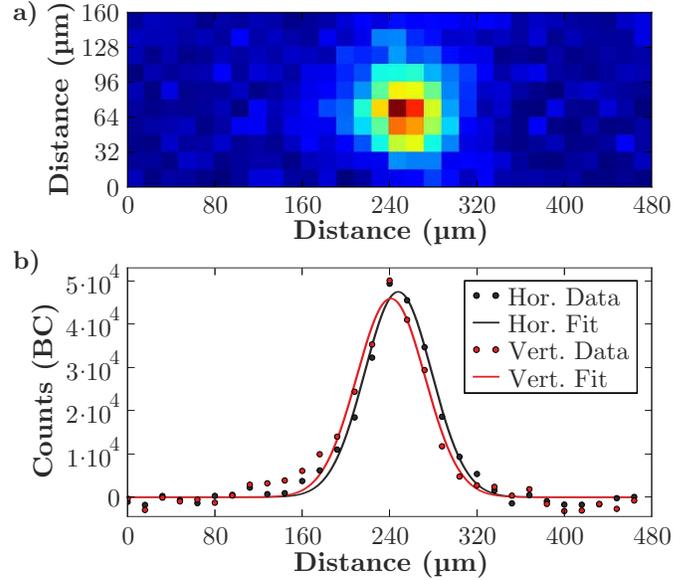}
	\end{center}
	\caption{Panel a) depicts an image of a $^{\mathrm{40}}$Ca$^{\mathrm{+}}$ ion on an EMCCD camera. Panel b) displays the
		averaged background corrected (BC) counts of the ion along a horizontal and vertical axis.  The counts represent 
		Gaussian fits showing few aberrations in the detection optics. }
	\label{fig: detected ion}
\end{figure}

We can use an ion image on the EMCCD camera to 
characterize the detection optics, as shown with a $^{\mathrm{40}}$Ca$^{\mathrm{+}}$ ion in 
Fig.~\ref{fig: detected ion}.  The magnification of the imaging system is 15, and the sum over the rows 
and columns of the region of interest
are in agreement with Gaussian fits of 1.84(7)~$\mathrm{\mu}$m width and 1.89(11)~$\mathrm{\mu}$m height.

\begin{figure}[!bp]
	\begin{center}
		\includegraphics[scale=0.3]{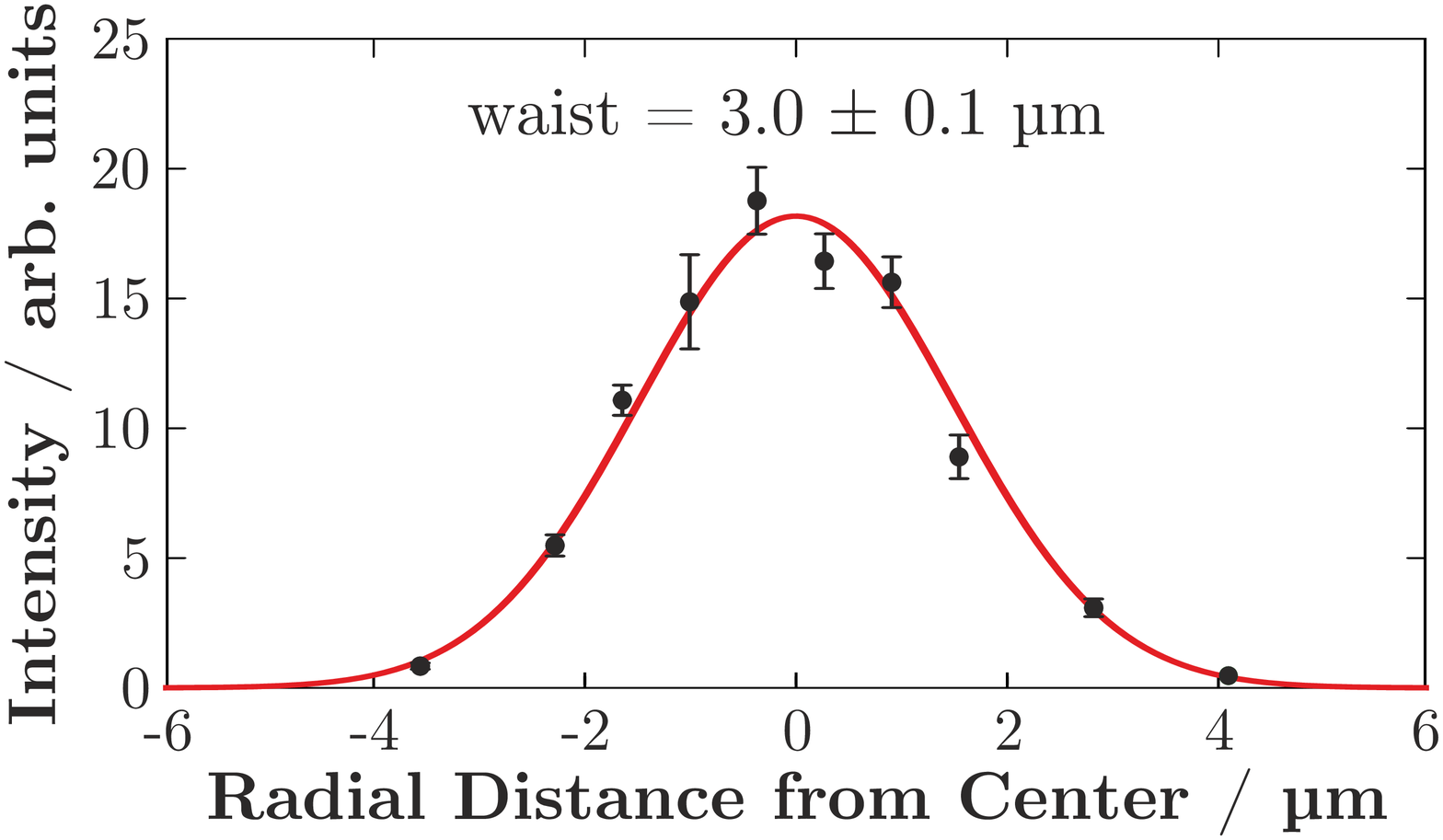}
	\end{center}
	\caption{Test of our addressing optics. A single ion was moved along the trap
		axis through the addressing beam and the measured light intensity is measured,
		resulting in a beam waist of 3.0(1)~$\mathrm{\mu}$m.}
	\label{fig: addressing}
\end{figure}

Next, we characterized the qubit addressing capabilities\cite{AddressingHCN} of the NA $= 0.23$ lens.
For this measurement, a $^{\mathrm{40}}$Ca$^{\mathrm{+}}$ ion is moved along the trap axis through the
addressing beam and the Rabi flops for different positions were measured. From the Rabi frequency, 
the light intensity at one position along the trap axis can be determined.
The obtained intensity profile of the beam along the trap axis is depicted in Fig.~\ref{fig: addressing}. 
The measured beam waist is 3.0(1)~$\mathrm{\mu}$m. 

\begin{figure}[!tp]
	\begin{center}
		\includegraphics[scale=0.3]{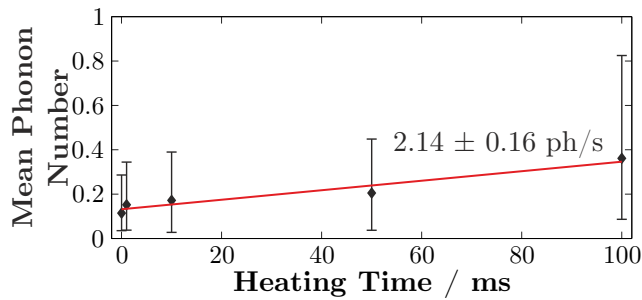}
	\end{center}
	\caption{Axial heating rate measurement for a trap with an ion-electrode distance of 
		113~$\mathrm{\mu}$m and a secular motion frequency of 1.1~MHz.}
	\label{fig: heating rate}
\end{figure}

Motional coherence is required in entangling gate operations\cite{CZ,MS-gate}. 
The heating rate is a rate of change of motional quanta of the secular 
motion in phonons per second, and high motional heating rates in a trap limit the 
fidelity of entangling gates.  Thus, the heating rate of a trap is a quantity of 
interest for trapped ion quantum information processing setups. 
The heating rate is obtained by measuring the mean phonon number
after cooling and idling for different wait times, as shown in Fig.~\ref{fig: heating rate}.  The measured heating
rate for the axial mode of vibration of a single ion is 2.14(16)~phonons/s, which is similar to results of other
experiments with an electrode-ion distance close to our distance of 113~$\mathrm{\mu}$m and a secular motion 
frequency close to 1.1~MHz in cryogenic setups\cite{heatingRatesMike}.
From this, we conclude that the slot in the center of the trap does not affect the heating rate.

\begin{figure}[!bp]
	\begin{center}
		\includegraphics[scale=0.3]{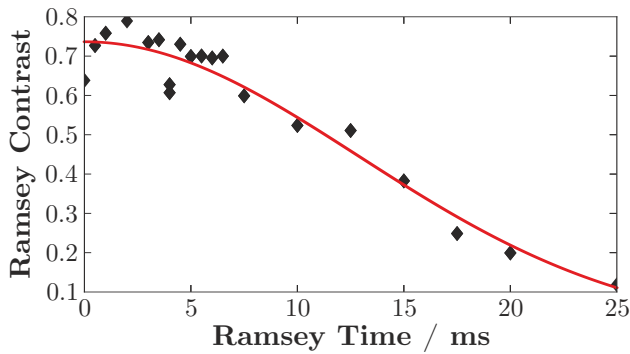}
	\end{center}
	\caption{Ramsey experiment with an optical $^{\mathrm{40}}$Ca$^{\mathrm{+}}$ qubit on 
		the $\left|S,m_\mathrm{j}=-1/2\right\rangle$ to $\left|D,m_\mathrm{j}=-1/2\right\rangle$
		transition in our setup, during which the optical table did not float, resulted in a Gaussian 
		decay with a \nicefrac{1}{e}-time of 18.2(8)~ms.}
	\label{fig: ramsey}
\end{figure}

Another important quantity to characterize the performance of an apparatus for quantum information processing is the coherence time.  
In a Ramsey experiments, we prepare a balanced
superposition of the two qubit states.  After a wait time, one tries to rotate the state
back into one of the two qubit states.  By varying the phase of the second operation, one obtains a sine-like dependence
of the excitation probability over the varied phase.  The contrast of this sine contains information
on the amount of noise present during the wait time.  Hence, the Ramsey contrast in dependence
of the wait time is a measure of coherence of the system.

The Ramsey measurement on the $\left|S,m_\mathrm{j}=-1/2\right\rangle$ to $\left|D,m_\mathrm{j}=-1/2\right\rangle$
transition, shown in Figure~\ref{fig: ramsey}, resulted in a Gaussian
decay of the Ramsey contrast with a \nicefrac{1}{e}-time of 18.2(8)~ms.
While the observed coherence time is sufficient for performing quantum gate operations, still, 
from the measured magnetic shielding efficiency of the inner heat shield, we would expect that 
the qubit phase coherence is ultimately limited by the qubit state lifetime. 
In future, we plan to float the optical table on its air-damped platform to minimize
the dependence of the coherence of an optical qubit on acoustic vibrations. 

\section{Conclusion}
\label{sec: conclusion}

In this work, we presented our new cryogenic ion trap experiment containing observation 
and addressing optics, and RF enhancement resonator inside inner heat shield.  
A detailed study of the mechanical vibrations was shown which yielded vibrations
on the order to 1~\% of the qubit laser wavelength.  Furthermore, a frequency 
analysis of the shielding against magnetic field noise in the lab was performed
resulting in an upper bound of -120~dB of the reduction of 50~Hz magnetic
signals along the quantization axis.

In addition to our cryostat, we described the laser setup required for the 
operation with $^{\mathrm{40}}$Ca$^{\mathrm{+}}$ and 
$^{\mathrm{88}}$Sr$^{\mathrm{+}}$ ions.  The qubit laser for
$^{\mathrm{40}}$Ca$^{\mathrm{+}}$ could be characterized with a second narrow 
laser at the same frequency.  We observed a beat frequency of 1.58(2)~Hz 
between the two lasers and a lowest Allan deviation of 
2.4$\cdot$10$^{\mathrm{-15}}$ at a time interval of 0.33~s.

Furthermore, we presented our new electronic setup, which allowed us to 
perform first measurements with ions.  The detection and addressing optics 
were characterized.  Moreover, Rabi flops, heating rates, and
coherence times were measured.  Because of the low vibrations, the high
magnetic shielding, suitable laser and electronics setup, we believe
this work provides an experimental platform suitable for
a fault-tolerant trapped ion quantum computer.

\begin{acknowledgments}
	This research was funded by the Office of the Director of National Intelligence (ODNI),
	Intelligence Advanced Research Projects Activity (IARPA), through the Army Research Office grants
	W911NF-10-1-0284 and W911NF-16-1-0070.  All statements of fact, opinion or conclusions contained herein are those of the
	authors and should not be construed as representing the official views or policies of IARPA, the ODNI,
	or the U.S. Government.  Financial support was provided by the Austrian Science Foundation (FWF), through 
	the SFB FoQuS (FWF project F4002-N16) as well as the Institut f\"ur Quantenoptik und Quanteninformation GmbH.
	P.S. was supported by the FWF Erwin Schr\"odinger Stipendium 3600-N27.
\end{acknowledgments}

\bibliography{Cryostina}

\end{document}